\documentclass[aps,prl,twocolumn,byrevtex,showpacs,preprintnumbers,amsmath,amssymb]{revtex4}

\usepackage{graphicx}
\usepackage{color}

\begin{document}
\title{Emerging beam resonances in atom diffraction from a
reflection grating}
\author{Bum Suk Zhao}
\email{zhao@fhi-berlin.mpg.de}
\author{Gerard Meijer}
\author{Wieland Sch\"ollkopf}
\email{wschoell@fhi-berlin.mpg.de} \affiliation{Fritz-Haber-Institut
der Max-Planck-Gesellschaft, Faradayweg 4-6, 14195 Berlin, Germany}
\date{\today}

\begin{abstract}
We report on the observation of emerging beam resonances, well known
as Rayleigh-Wood anomalies and threshold resonances in photon and
electron diffraction, respectively, in an atom-optical diffraction
experiment. Diffraction of He atom beams reflected from a blazed
ruled grating at grazing incidence has been investigated. The total
reflectivity of the grating as well as the intensities of the
diffracted beams reveal anomalies at the Rayleigh angles of
incidence, i.e., when another diffracted beam emerges parallel to
the grating surface. The observed anomalies are discussed in terms
of the classical wave-optical model of Rayleigh and Fano.
\end{abstract}


\maketitle

It is a general phenomenon in wave optics that diffraction by a
periodic surface shows peculiar intensity variations in the outgoing
beams when conditions (i.e.\ wavelength, periodicity, and incidence
angle) are such that a diffracted beam just emerges parallel to the
surface. This was first observed by Wood in 1902 \cite{Wood02} who
found strange dark and bright bands (Wood anomalies) in diffraction
patterns of white light from ruled reflection gratings. Rayleigh in
1907 first traced back Wood anomalies to grazing emergence of a
diffracted beam \cite{Rayleigh1907a}. Thus, the term Rayleigh
conditions (Rayleigh wavelength, Rayleigh angle) refers to
conditions for grazing beam emergence where Wood anomalies occur.
Rayleigh and subsequently Fano explained the anomalous behavior by
interference between first and second order scattering of the
incident wave, where second order scattering refers to waves that
are first scattered from a neighboring grating stripe
\cite{Rayleigh1907a,Fano38,Fano41}. Later on Wood anomalies were
categorized into two cases, one (sharp anomalies) related
exclusively to the emergence of a diffracted beam (Rayleigh-Wood
anomaly) and the other (broad anomalies) to resonance effects
\cite{Fano41,Hessel65}. The resonance type Wood anomaly is
attributed to excitation of surface-plasma oscillations guided along
the grating surface \cite{Teng67,Ritchie68}. In recent years the
effect of extraordinary optical transmission through periodic arrays
of sub-wavelength holes \cite{Ebbesen98} has been explained in terms
of Wood anomalies \cite{Sarrazin2003,Garcia2007}. In addition,
Rayleigh-Wood anomalies have been observed in soft x-ray diffraction
\cite{Fujizara68} and have been considered in designing x-ray
monochromators \cite{Padmore94}.

The effect has also been studied independently in matter-wave optics
such as reflection high energy electron diffraction (RHEED), low
energy electron diffraction (LEED), and atom and molecular beam
scattering from crystal surfaces \cite{McRae79}. In LEED, for
example, it was first observed in unfocused electron beam
diffraction from crystal surfaces \cite{McRae67,McRae71}. Here, a
crystal surface was used instead of a grating surface, since a
periodic length on the order of an $\rm \AA$ is required for the
diffraction of the electron due to its small de Broglie wavelength.
The parlance used in electron diffraction is, however, different
from classical optics using the terms {\it threshold effect} and
{\it electronic surface resonance} instead of the counterparts in
optics, Rayleigh-Wood and resonance type anomaly, respectively
\cite{McRae79}. Also, the emerging beam condition given by the
Rayleigh wavelength was referred to as type K$_{\rm II}$ Kikuchi
lines \cite{Somorjai69}.

In atom optics a behavior in analogy to the resonance type anomaly
and the electronic surface resonance has been investigated for a
long time under the name of selective adsorption \cite{Hulpke92sa}.
Estermann and Stern in 1930 observed anomalous intensity
fluctuations in the specular peak of helium diffraction from a
crystal surface \cite{Estermann30}. The anomaly was accounted for by
a bound state of the atom-surface interaction potential
\cite{Lennard36}. The relation between selective adsorption and Wood
anomaly was discussed by Fano already in 1938 \cite{Fano38}. On the
other hand, Rayleigh-Wood anomalies have not been observed in atom
surface scattering experiments, although they have been predicted by
theory \cite{Cabrera74,Garcia82,Armand86}. In these theoretical
studies the anomaly was referred to as {\it threshold resonance} or
{\it emerging beam resonance}. As the latter term is more
descriptive, it is adopted here. More recently, Guantes {\it et
al.}\ suggested that emerging beam resonances should be observable
in elastic atom scattering from a highly corrugated surface
\cite{Guantes97}.

Here we report the first observation of emerging beam resonances in
an atom surface scattering experiment, complementing Rayleigh-Wood
anomalies observed with light and electrons. Helium atom beams are
diffracted from a plane ruled grating at near grazing incidence. By
varying the incidence angle we observe the resonances precisely at
the Rayleigh incidence angles in two ways; (i) the total coherent
reflectivity of the grating increases steeply, (ii) the intensity
curves of the diffracted beams and the specular beam exhibit abrupt
changes of their slopes. We adopt the multiple scattering approach
of Rayleigh \cite{Rayleigh1907a} and Fano \cite{Fano38,Fano41} to
explain basic features of our observations.

As predicted by Armand and Manson \cite{Armand86} emerging beam
resonances in atom surface scattering occur within a small angular
range. Thus, the collimation of the incident beam and the angular
resolution of detection are required to be on the order of 100
$\mu$rad. Another prerequisite for observing the effect is to have a
significant flux diffracted into the very beam that, at Rayleigh
conditions, emerges above the surface so as to have an appreciable
effect on the other outgoing beam intensities. In the experiment
described here these requirements have been met by reflecting a
highly collimated helium atom beam at grazing incidence from a
blazed ruled diffraction grating. The use of helium atoms at grazing
incidence ensures sufficient coherent reflection probability
\cite{Zhao08}, while the grating blaze angle in combination with a
well chosen azimuthal orientation of the grating leads to an
effective enhancement of the intensity of the emerging diffracted
beam.

% ---------------------------------     Setup  -------------------------------

%%------------------------ Fig 1 ----------------------------
\begin{figure}[b]
\includegraphics[scale=0.32]{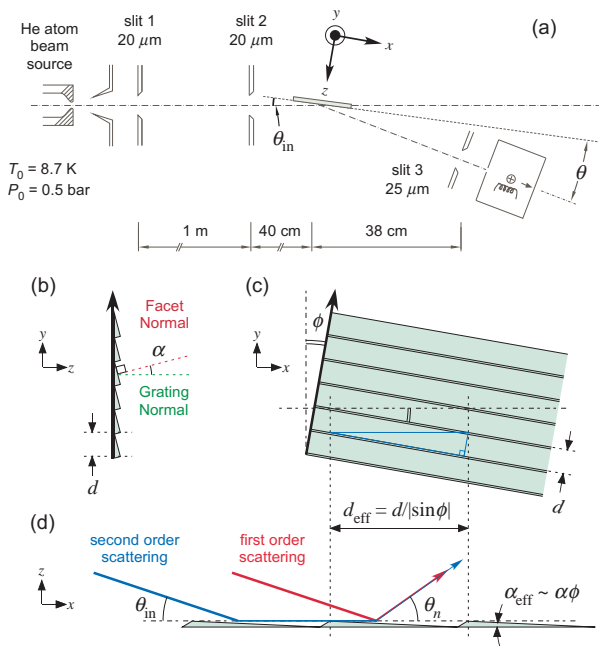}
\caption{(Color online) Scheme of the experimental setup and
orientation of the plane ruled grating. In each figure the chosen
coordinate system is denoted. The grating azimuth angle $\phi$ is
the angle between the grating blaze arrow (thick arrow in (b) and
(c)) and the $y$-axis. In (b) $\phi = 0$, whereas (c) and (d)
correspond to $\phi < 0$.} \label{fig:setup}
\end{figure}

The high angular resolution diffraction apparatus has been used in
previous experiments \cite{Zhao08,Schewe09}. The continuous atom
beam is formed by supersonic expansion of He gas at stagnation
temperature $T_0 = 8.7$ K and pressure $P_0 = 0.5$ bar through a
5-$\mu$m-diameter orifice into high vacuum. After passing a skimmer
of 500 $\mu$m diameter, the beam is collimated by two 20 $\mu$m wide
slits (slit~1 and slit~2) separated by 100 cm as indicated in
Fig.~\ref{fig:setup}. In combination with the 25 $\mu$m wide
detector-entrance slit (slit~3), located 78 cm downstream from the
second slit, the observed angular width of the atom beam is 120
$\mu$rad (full width at half maximum). The third slit and the
detector (an electron-impact ionization mass spectrometer) are
mounted on a frame which is rotated precisely as indicated in
Fig.~\ref{fig:setup}(a). A plane ruled grating is positioned such
that the detector pivot axis is parallel to the grating surface and
passes through its center. The pivot axis (vertical), and the
grating normal (horizontal), are chosen as the $y$- and $z$-axis of
the reference frame, respectively. Hence, the (horizontal)
$xz$-plane is the plane of incidence. The grazing incidence angle
$\theta_{\rm in}$ and the detection angle $\theta$ are measured with
respect to the grating surface plane. Diffraction patterns are
obtained by rotating the detector, namely varying $\theta$, and
measuring the He signal at each angle.

The commercial plane ruled grating (Newport 20RG050-600-1) is made
out of 6 mm thick glass with an aluminum coating and has a surface
area of 5 x 5 cm$^2$. It is characterized by a period $d = 20$
$\mu$m and a blaze angle $\alpha = 0.8^{\circ}$ ($\approx 14$ mrad)
(see Fig.~\ref{fig:setup}(b)). The orientation of the grating in
space is defined by the grating normal and the blaze arrow. The
latter is perpendicular to the grating normal and to the grating
grooves and points vertically upward in Fig.~\ref{fig:setup}(b).

The grating is aligned such that its grooves are almost (but not
quite) parallel to the $x$-axis (horizontal). This geometry
corresponds to the conical diffraction mode known from EUV
spectroscopy \cite{Cash82}. In this geometry out-of-plane
diffraction is without effect on the measurements because the
vertical acceptance angle of slit 3 ($\approx 10$ mrad) is far
larger than the vertical diffraction angles (tens of $\mu$rad). As
indicated in Fig.~\ref{fig:setup}(c) the grating can be rotated by
the azimuth angle $\phi$ around the $z$-axis. By varying $\phi$ the
effective periodic length $d_{\rm eff}$ and the effective blaze
angle $\alpha_{\rm eff}$ can be adjusted, as depicted in
Figs.~\ref{fig:setup}(c) and (d). The former is given by
trigonometry of the triangle in Fig.~\ref{fig:setup}(c), $d_{\rm
eff}=d/|\sin\phi|$, while the latter is approximated by $\alpha_{\rm
eff} \approx \alpha \phi$, which is derived from $\sin\alpha_{\rm
eff} = \sin\alpha \sin\phi$. We define $\phi$ to be negative
(positive) when the blaze arrow is rotated clockwise
(counterclockwise) from the $y$-axis. In this convention
Fig.~\ref{fig:setup}(c) shows the case of $\phi<0$. Since the effect
of blazing is to enhance those diffracted beams which are specularly
reflected with respect to the facet normal, negative (positive)
diffraction orders get enhanced for $\phi<0$ ($\phi>0$).

% -----------------------------------      Results   -----------------------------------

%%------------------------ Fig 2 total reflectivity  --------------------------------
\begin{figure}[bth]
\includegraphics[scale=0.53]{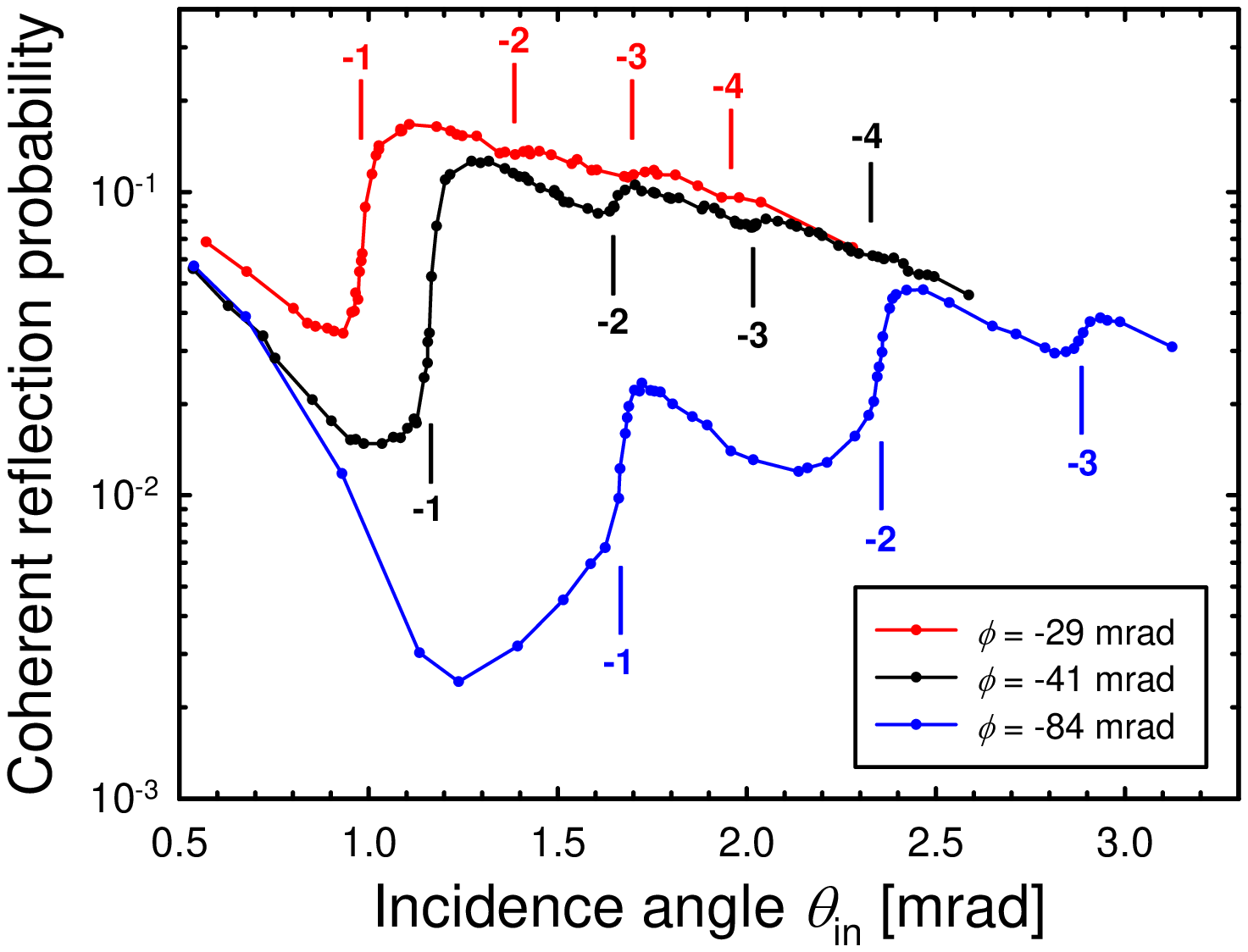}
\caption{(Color online) Fraction of He atoms that are coherently
scattered from the plane ruled grating for different azimuth angles
$\phi = -29$, $-41$, $-84$~mrad. Vertical lines indicate the
positions of the Rayleigh angles of incidence with the numbers
indicating the diffraction order of the emerging beam. (The
diffraction order sign convention follows Ref.~\cite{Padmore94}.)}
\label{fig:reflectivity}
\end{figure}

The fraction of He atoms that are coherently scattered from the
grating, measured for three different azimuth angles $\phi<0$, is
plotted as a function of incidence angle in
Fig.~\ref{fig:reflectivity}. It is determined from the summation of
the areas $A_{n}$ of the diffraction peaks in intensity vs.\
detection angle plots (see below). The sum is taken over all
diffraction orders $n$ (including $n=0$, the specular peak) and
normalized to the incidence beam area $A_{\rm in}$. The latter is
measured when the grating is moved out of the beam path. The
vertical lines in Fig.~\ref{fig:reflectivity}, each labeled by an
integer, indicate the positions of the Rayleigh angles of incidence
$\theta_{{\rm R},m}$ with the integer indicating the diffraction
order $m$ of the emerging beam. The Rayleigh angle of incidence is
calculated from the grating equation, $\cos{\theta_{{\rm R},m}}-1 =
m\frac{\lambda}{d_{\rm eff}}$ \footnote{For the conical diffraction
geometry a modified grating equation holds, taking into account the
out-of-plane (i.e.\ conical) diffraction \cite{Cash82}. However, for
our conditions (i.e.\ $|n\lambda/d| \ll |\sin{\phi}|$) the in-plane
grating equation employed with $d_{\rm eff}$ is a very good
approximation.}. ($\lambda = 3.32$~\AA\ is the de Broglie wavelength
of the helium atoms at $T_0=8.7$ K).

The coherent reflection probability curves are not monotonically
decaying with incidence angle as observed in previous diffraction
experiments \cite{Zhao08}. Instead, strong and abrupt variations are
found at exactly the Rayleigh angles. This indicates that emergence
of another diffraction beam not only leads to a redistribution of
flux among the outgoing beams, but also to an increase of the
fraction of He atoms that are coherently scattered. There must be a
concurrent decrease of the fraction of He atoms that undergo diffuse
scattering at the surface.

\begin{figure}[bth]
\includegraphics[scale=0.53]{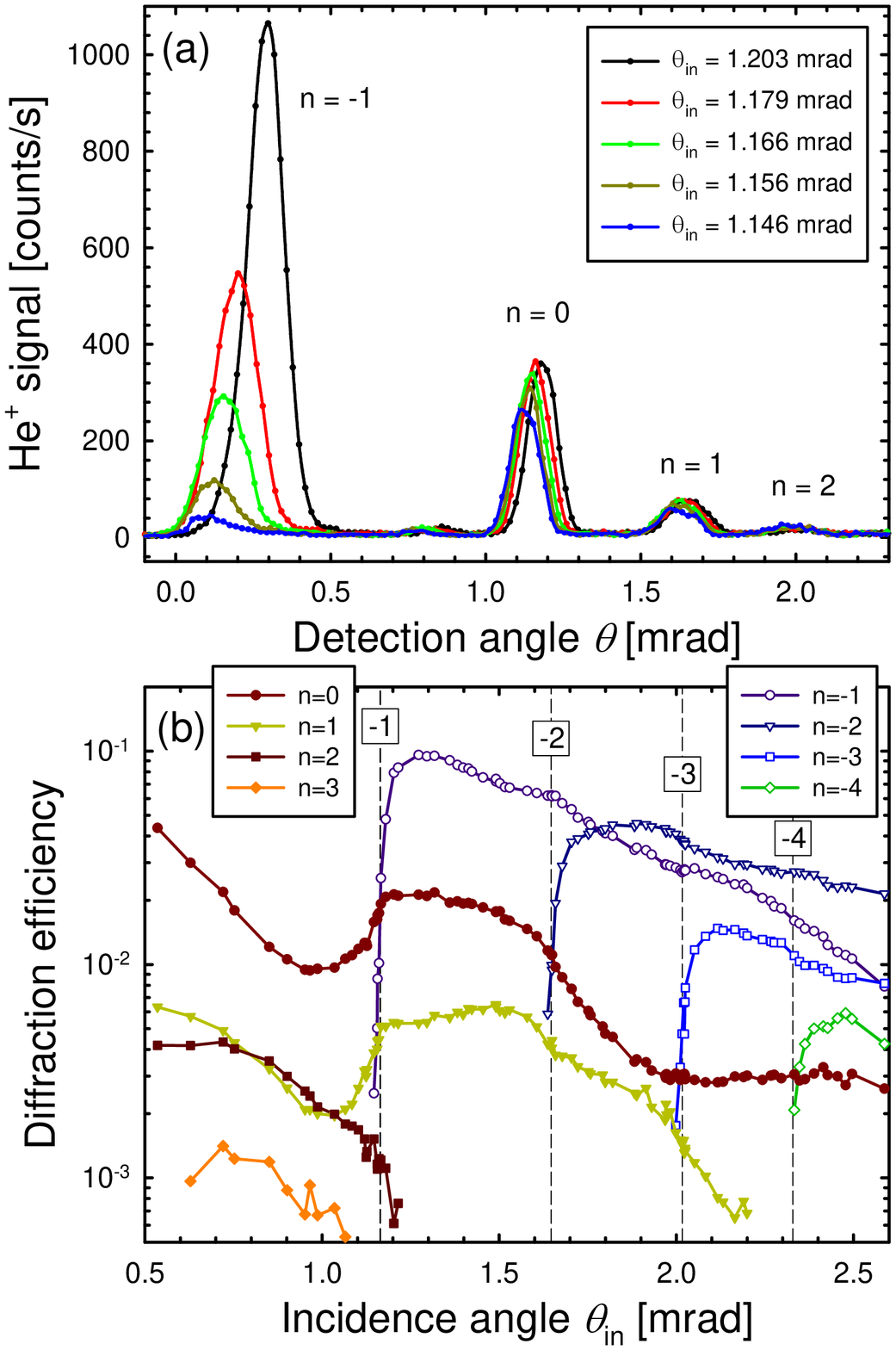}
\caption{(Color online) (a) Diffraction patterns of He atom beams
for $\phi = -41$~mrad for 5 different incidence angles in the
vicinity of the Rayleigh incidence angle $\theta_{{\rm R},-1}=1.164$
mrad where the $-1^{\rm st}$ order peak emerges, (b) diffraction
efficiencies for $\phi = -41$~mrad as a function of incidence angle.
The dashed vertical lines indicate the calculated Rayleigh incidence
angles corresponding to the emergence of the $-1^{\rm st}$, $-2^{\rm
nd}$, $-3^{\rm rd}$, and $-4^{\rm th}$ order diffraction peaks.}
\label{fig:SpeDE}
\end{figure}

The azimuth angle $\phi$ is obtained by analyzing diffraction
patterns at various incidence angles. Fig.~\ref{fig:SpeDE}(a) shows
diffraction spectra for $\phi = -41$~mrad at five different
incidence angles in the vicinity of the Rayleigh incidence angle
$\theta_{{\rm R},-1}=1.164$ mrad, where the $-1^{\rm st}$ order peak
emerges. The diffraction angles $\theta_{n}$ and areas $A_{n}$ of
the $n^{\rm th}$-order diffraction peak are found by fitting each
peak with a Gaussian curve. The incidence angle $\theta_{\rm in}$ is
determined from the detection angle of the specular peak,
$\theta_{0}$. The diffraction angles $\theta_{n}$ as a function of
$\theta_{\rm in}$ are then fitted by the grating equation,
$\cos{\theta_{\rm in}}-\cos{\theta_n}= n\frac{\lambda}{d_{\rm eff}}$
with $d_{\rm eff}$ being the only fit parameter. For the data set in
Fig.~\ref{fig:SpeDE} $d_{\rm eff} = 493 \pm 1$ $\mu$m is found
corresponding to an azimuth angle $\phi=-41$ mrad.

The angular spectra shown in Fig.~\ref{fig:SpeDE}(a) illustrate the
progressive emergence of the $-1^{\rm st}$-order peak. The height of
the latter increases rapidly from 40 to 1050 counts/s within this
range of incidence angles, whereas the height of the specular peak
increases from 270 to 360 counts/s when $\theta_{\rm in}$ is
increased from 1.146 to 1.179 mrad, and stays around 360 counts/s at
$\theta_{\rm in}=1.203$ mrad. This indicates a discontinuity of the
slope of the specular intensity variation around $\theta_{\rm
in}=1.179$ mrad, which agrees with the calculated Rayleigh angle. As
can be seen in Fig.~\ref{fig:SpeDE}(a) this coincides with the
incidence angle at which the $-1^{\rm st}$-order diffraction peak
starts to be fully separated from the surface. The high intensity of
the diffraction beam at grazing emergence is remarkable and might be
a manifestation of the reciprocity theorem, as observed before in
x-ray grating diffraction \cite{Jark88}.

%%------------------------ Fig 3 angular profiles ----------------------------

Details of the emerging beam resonances can be seen more easily in
the semi-logarithmic plot of Fig.~\ref{fig:SpeDE}(b) showing the
diffraction efficiency $A_n/A_{\rm in}$ over a large range of
incidence angles from 0.5 to 2.6 mrad. The different symbols
correspond to various diffraction orders from $n=-4$ to 3. The
dashed vertical lines indicate the Rayleigh angles of incidence.
Figure~\ref{fig:SpeDE}(b) exemplifies general aspects of emerging
beam resonances, which have also been found for further azimuth
angles not shown here. (I) The resonance behavior at $\theta_{{\rm
R},m}$ observed in the outgoing beams of order $n$ (with $n \neq m$)
is the more distinctive, the more intense the emerging beam is. This
can be seen in the figure where the resonance behavior is most
pronounced at $\theta_{{\rm R},-1}$, less distinct at $\theta_{{\rm
R},-2}$, and hardly visible when the less intense $-3^{\rm rd}$ and
$-4^{\rm th}$-order diffraction peaks emerge. (II) Discontinuities
seem to appear in all the other diffraction peaks, although they are
most pronounced in neighboring diffraction orders, namely, $A_{m+1}$
and $A_{m+2}$. In Fig.\ \ref{fig:SpeDE}(b), for instance, at
$\theta_{{\rm R},-1}$ pronounced discontinuities are found in the
slopes of $A_0$ and $A_1$. (III) At the Rayleigh angle of incidence
the slope of $A_n$ usually exhibits a discontinuous decrease with
increasing incidence angle, except for a few cases where a
discontinuous increase is observed. In the figure, $A_0$ and $A_1$
are found to increase steeply (with the slope being close to
diverging) at $\theta_{\rm in} \leq \theta_{{\rm R},-1}$, whereas
they hardly change right after the vertical line. Similarly, at
$\theta_{{\rm R},-2}$ the slope of $A_{-1}$ shows a sudden decrease,
the slope of $A_1$, however, abruptly increases.

Following the approach introduced by Rayleigh and Fano
\cite{Rayleigh1907a,Fano38,Fano41} the amplitude of the $n^{\rm
th}$-order diffraction $S_n$ is approximated by interference of the
amplitudes from the first and the second order scattering, $S_n =
S^{(1)}_n + S^{(2)}_n$. The first order scattering is the scattering
of the incident beam at a given grating unit (red arrow in
Fig.~\ref{fig:setup}(d)), while the second order scattering is the
scattering of a beam at that grating unit, which has already
undergone scattering at another grating unit (blue arrow in
Fig.~\ref{fig:setup}(d)). When the $m^{\rm th}$-diffraction-order
fulfills the Rayleigh condition, $S^{(2)}_n$ is proportional to
$S^{(1)}_m$ and the interference is constructive. Thus, the sudden
increase of the emerging $m^{\rm th}$-order intensity increases the
other orders (aspects (II) and (III)). The degree of the influence
is proportional to $S^{(1)}_m$, namely, $A_m$ (aspect (I)).
Therefore, once the $m^{\rm th}$-order peak gets separated from the
grating, as seen in the spectra of Fig.~\ref{fig:SpeDE}(a), the
contribution of the second order scattering to the other diffraction
beams diminishes and their intensities level off.

Emerging beam resonances, i.e., Rayleigh-Wood anomalies in atom
optics, have been predicted to provide detailed information of
atom-surface interaction potentials
\cite{Cabrera74,Garcia82,Armand86}. However, more than a century
after the first observation of Wood anomalies \cite{Wood02} and 80
years after the first observation of selective adsorption, i.e., a
resonance type Wood anomaly in atom surface scattering
\cite{Estermann30}, they were still not observed in atom diffraction
experiments. Here we report the first observation of emerging beam
resonances in atom optics using a blazed ruled grating at grazing
incidence with the grooves oriented almost parallel to the incident
beam direction. The total coherent reflectivity of the grating as
well as the intensities of the diffracted beams reveal anomalies at
the Rayleigh angles of incidence which are interpreted with the
approach developed many decades ago to describe the anomalies
observed with photons \cite{Rayleigh1907a,Fano41}. Therefore, this
observation completes the analogy between photon optics ({\it
Rayleigh-Wood anomaly} and {\it resonance type Wood anomaly}) on the
one hand and atom optics ({\it emerging beam resonance} and {\it
selective adsorption}) on the other hand.

B.S.Z.\ acknowledges support by the Alexander von Humboldt
Foundation and by the Korea Research Foundation Grant funded by the
Korean Government (KRF-2005-214-C00188). We thank J.R.\ Manson for
insightful discussions and H.C.\ Schewe for help with the apparatus.

\end{document}